\documentclass[prl,twocolumn,showpacs]{revtex4-1}

\usepackage{graphicx}
\usepackage{amsmath}
\usepackage{amssymb}
\usepackage{epsfig}
\usepackage{subfigure}

\renewcommand{\v}[1]{{\bf #1}}

\def\eqa{\begin{eqnarray}}
\def\eea{\end{eqnarray}}
\newcommand{\eq}{\begin{equation}}
\newcommand{\ee}{\end{equation}}

\newcommand{\ua}{\uparrow}
\newcommand{\da}{\downarrow}
\newcommand{\ra}{\rightarrow}

\newcommand{\al}{\alpha}
\newcommand{\bt}{\beta}

\newcommand{\ga}{\gamma}

\newcommand{\La}{\Lambda}

\newcommand{\si}{\sigma}

\newcommand{\vk}{{\bf k}}

\usepackage{color}

\begin{document}

\title{Functional renormalization group study of the pairing symmetry and pairing mechanism in iron-selenide superconductors}

\author{Yuan-Yuan Xiang}
\affiliation{National Laboratory of Solid State Microstructures,
Nanjing University, Nanjing, 210093, China}

\author{Yang Yang}
\affiliation{National Laboratory of Solid State Microstructures,
Nanjing University, Nanjing, 210093, China}

\author{Wan-Sheng Wang}
\affiliation{National Laboratory of Solid State Microstructures,
Nanjing University, Nanjing, 210093, China}

\author{Zheng-Zao Li}
\affiliation{National Laboratory of Solid State Microstructures,
Nanjing University, Nanjing, 210093, China}

\author{Qiang-Hua Wang}
\email{qhwang@nju.edu.cn}\affiliation{National Laboratory of Solid
State Microstructures, Nanjing University, Nanjing, 210093, China}

\begin{abstract}
In iron selenide superconductors only electron-like Fermi pockets survive, challenging the $S^{\pm}$ pairing based on the quasi-nesting between the electron and hole Fermi pockets (as in iron arsenides). By functional renormalization group study we show that an in-phase $S$-wave pairing on the electron pockets ($S^{++}_{ee}$) is realized. The pairing mechanism involves two competing driving forces: The strong C-type spin fluctuations cause attractive pair scattering between and within electron pockets via Cooperon excitations on the virtual hole pockets, while the G-type spin fluctuations cause repulsive pair scattering. The latter effect is however weakened by the hybridization splitting of the electron pockets. The resulting $S^{++}_{ee}$-wave pairing symmetry is consistent with experiments. We further propose that the quasiparticle interference pattern in scanning tunneling microscopy and the Andreev reflection in out-of-plane contact tunneling are efficient probes of in-phase versus anti-phase $S$-wave pairing on the electron pockets.
\end{abstract}


\maketitle

Soon after the discovery of iron pnictide superconductors~\cite{Hosono1111}, it was realized that the quasi-nesting between the electron and hole pockets enhances spin fluctuations at the nesting vectors. In the parent compound the spin fluctuations are so strong that a co-linear C-type spin-density-wave (SDW) state develops~\cite{c_sdw}. By carrier doping or physical/chemical pressure the spin fluctuations are weakened, but they act as the seed for repulsive pair scattering between the electron and hole Fermi pockets, leading to anti-phase $S$-wave pairing on the electron and hole pockets (denoted as $S^{\pm}_{eh}$ henceforth)~\cite{s+-}. Such a pairing symmetry turns out to be consistent with the neutron resonance mode in the inelastic neutron scattering~\cite{INS122} and the quasi-particle scattering interference pattern in scanning tunneling microscopy measurements~\cite{HanagurySTM}.

However, the $S^{\pm}_{eh}$-scenario is questioned for iron-selenides.~\cite{KFS1,KFS2}. Initially the $\sqrt{5}\times\sqrt{5}$ vacancy ordered phase was conjectured as the parent phase of superconductivity,~\cite{bao,weng} but it is most likely phase separated from the true superconducting phase given the fact that the transition temperature hardly changes against significant variations in the nominal doping.~\cite{KFS1,KFS2} In contrast to the iron pnictides, here the hole-like bands sink below the Fermi level~\cite{KFSFS}, banning low energy scattering between electron and hole pockets. The low energy spin scattering between the remaining electron Fermi pockets (not nested in a strict sense since both are electron-like) would favor a nodeless $d$-wave pairing, with opposite gap signs on the electron pockets~\cite{KFSD,Thomale,millis}. However, from a three-dimensional view of the Fermi surface, the intersection of electron pockets (henceforth in the reduced small Brillouine zone unless specified otherwise) varies in $k_z$ and would force the $d$-wave pairing to generate nodes and thus become less favorable energetically.~\cite{KFSHybrid} From angle-resolved photoemission spectroscopy (ARPES) measurements the gaps are nodeless not only on electron pockets in the $k_z=0$ plane, but also on an electron pocket encircling the momentum $(0,0,\pi)$,~\cite{HDingKFeSeEPL,DLFengKFeSePRB} ruling out any form of $d$-wave pairing. There remains two possibilities for the full pairing gaps observed experimentally. On one hand, Mazin argued that a new type of pairing with anti-phase gaps on the hybridization split electrons pockets (denoted as $S^{\pm}_{ee}$ henceforth) would be favorable~\cite{KFSHybrid}. This scenario is pushed further by a phenomenological Ginzburg-Landau theory.~\cite{KFSS+-} On the other hand, theories based on orbital fluctuations~\cite{kontani} or a strong phenomenological spin exchange $J_2$ on second-neighbor bonds ~\cite{Thomale} end up with in-phase pairing gaps on the electron pockets ($S^{++}_{ee}$ henceforth). Given the discrepancy, a microscopic theory accounting for both the hybridization splitting and the effective spin-exchange coupling is called for.

The experimental signature for the inphase/antiphase pairing gaps is limited. In the $S^{\pm}_{ee}$ scenario, the spin scattering vector is close to the umklapp vector $\v G$, implying a neutron resonance near this vector. It corresponds to G-type checkerboard SDW fluctuations. However, experimentally the neutron resonance is halfway between $\v G$ and the C-type SDW vector $\v C$.~\cite{KFSINSInosovPRB,KFSINSInosovPRL,KFSINSInosovEPL}. The deviation to $\v G$ can only be resolved by considering the microscopic shapes and sizes of the electron pockets. On the other hand, there are also reports of neutron resonance ~\cite{KFSINSDaiPRB} or neutron diffraction pattern ~\cite{KFSINSBirgeneauPRL} at the wavevector $\v C$, exactly as in iron pnictides. The discrepancy between experiments requires a better microscopic understanding of the material.

We are thus motivated to perform a microscopic study for the pairing mechanism in iron-selenides. As discussed above, one of the key factors is the hybridization between the otherwise independent electron pockets. This requires us to take at least a double-layer model, a situation not yet addressed microscopically so far. On the other hand, in order to treat the spin, charge and pairing channels on equal footing, and in particular, the competing C-type and G-type spin fluctuations and their overlaps to the pairing interactions, functional renormalization group~\cite{wetterich,salmhoferReview,wangfaFRG} is one of the indispensable machineries. In this Letter we use the recently developed singular-mode functional renormalization group (SMFRG) method, which has been successfully applied in the contexts of iron-based superconductors and candidate models with correlation driven topological insulators/superconductors~\cite{SMFRGGraphene,SMFRGFS,SMFRGTS,SMFRGKagome}.

The main results of this Letter are as follows. By SMFRG study of a microscopic double-layer model for iron selenides, we find that the gap functions are in-phase on the electron pockets, namely an $S^{++}_{ee}$-wave pairing is realized. The underlying mechanism involves two competing driving forces. The strong C-type spin fluctuations cause attractive pairing scattering via Cooperon excitations on the virtual hole pockets, while the G-type checkerboard spin fluctuations cause repulsive pair scattering. The latter effect is weakened by hybridization splitting of the electron pockets. The resulting $S^{++}_{ee}$-wave pairing symmetry is discussed in view of experimental perspectives.\\

\begin{figure}
\includegraphics[width=0.5\textwidth]{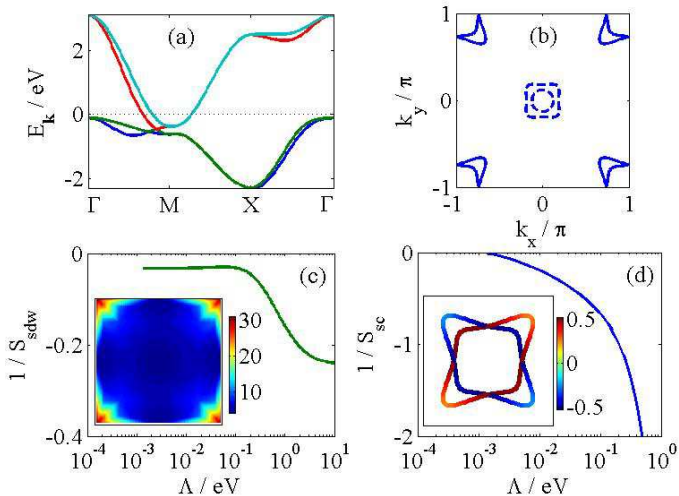}
\caption{The results for $t_{\perp}=0$. (a) Band structure along high symmetry cuts. (b) The electron-like Fermi pockets (solid lines) and the sunk hole pockets (dashed lines). The Fermi arcs in the four corners can be rearranged into closed pockets as shown around the right-top corner. (c) FRG flow of $1/S_{\rm SDW}$. The inset shows $S_{\rm SDW}(\v q)$ in the reduced Brillouine zone at the final stage. (d) FRG flow of $1/S_{SC}$. The inset shows the final $\phi_{SC}$ (color scale) projected on the Fermi surface.}\label{caseA}
\end{figure}

In our double-layer model, each layer contains two sublattices A and B, and they are connected vertically in an A-B fashion with an inter-layer hopping $t_{\perp}$, which hybridizes and splits the electron pockets. As advocated in Ref.\cite{HujpS4} it suffices to consider $(xz,yz)$ orbitals on each atom to capture the essential physics. For definiteness, we label a Fermion field at momentum $\v k$ by $c^{lmn}_\v k$, where $l=u/d$ labels the upper/lower layer, $m=A/B$ the sublattice, and $n=xz/yz$ the orbital. These fields can be assembled into an eight-component spinor field $\psi_\v k=(\{c^{lmn}_\v k\})^T$, so that the tight-binding hamiltonian can be written as, \eqa H_0=\sum_{\vk\si}\psi^\dag_{\v k\si}(h_{\vk}+t_{\perp}\tau_0s_1\ga_1)\psi_{\v k\si},\eea where $\si$ labels spin, $\tau_0$ is a unit matrix acting on orbitals, $s$ and $\ga$ are Pauli matrices acting on sublattice and layer bases, and finally $h_{\vk}$ is the intra-plane component that can be obtained from Ref.\cite{HujpS4}. The band structure for $t_{\perp}=0$ is shown in Fig.\ref{caseA}(a). Fig.\ref{caseA}(b) shows the Fermi surfaces (solid line) and the sunk hole pockets (dashed line). The band structure is similar for $t_{\perp}\neq 0$ (thus will not be reproduced), except that the electron pockets will be split (see below). In the following we tune $t_\perp$ systematically to see the effect of interlayer hybridization.

For the $d$-orbital system under concern, it is sufficient to consider local interactions,  $H_I=U\sum_{ia}n_{ia\ua}n_{ia\da}+U'\sum_{i,a>b}n_{ia}n_{ib}
+ J\sum_{i,a>b,\si,\si'}\psi_{ia\si}^\dag\psi_{ib\si}\psi_{ib\si'}^\dag
\psi_{ia\si'}+J'\sum_{i,a,b}\psi_{ia\ua}^\dag \psi_{ia\da}^\dag\psi_{ib\da}\psi_{ib\ua}.$
Here $i$ is a lattice site (on either sublattice and layer), $\si$ the spin, and $a$ and $b$ the orbital labels,
with $\psi_{a=1,2}$ annihilating an electron in $d_{xz}$ and $d_{yz}$
orbitals, respectively. As usual we use the Kanamori relations $U=U'+2J$ and $J=J'$ so that $(U,J)$ are the only two independent interaction parameters. In this paper $(U,J)$ are fixed at $(3.30,0.825)$eV. The interactions can lead to competing collective fluctuations in density-wave and pairing channels, which we handle by SMFRG as follows. A general interaction vertex
is decomposed as, \eqa V^{12;34}(\v k,\v k',\v q)\ra \sum_m
S_m (\v q) \phi_m^{12}(\v k,\v q)[\phi^{34}_m(\v k',\v q)]^*,\nonumber\eea
either in the superconducting (SC), spin density wave (SDW) or charge density wave (CDW) channels. Here the numbers are dummy labels for layer, sublattice and orbital indices, and $\v q$ is the collective momentum, $\v k$ (or $\v k'$) is an internal momentum of the Fermion bilinears $\psi^\dag_{\v k+\v q,1}\psi^\dag_{-\v k,2}$ and
$\psi^\dag_{\v k+\v q,1}\psi_{\v k,2}$ in the particle-particle
and particle-hole channels, respectively. In the following we define, in a specific channel,  $S(\v q)$ as the leading attractive eigenvalue at $\v q$, and $S$ the globally leading one. The SMFRG provides the {\em coupled} flow of all channels versus a decreasing energy scale $\La$ (the infrared limit of the Matsubara frequency in our case). The fastest growing eigenvalue implies an emergent order associated with an ordering wave vector $\v Q$ and a form factor $\phi(\v k,\v Q)$. (Notice that $\v Q=0$ in the SC channel because of the Cooper instability, but may evolve with $\La$ in the other channels.) The divergence energy scale is an estimate for the ordering temperature. More technical details can be found elsewhere~\cite{SMFRGGraphene,SMFRGFS,SMFRGTS}.

We begin with $t_\perp=0$. Since the layers are decoupled in this case, the actual calculation is performed for one-layer only. Since $|S_{\rm CDW}|$ remains small at low energy scales we shall not address it henceforth. The flow of $S_{\rm SDW}$ is shown in Fig.\ref{caseA}(c). It is enhanced in the intermediate stage and levels off at lower energy scales. Here the associated wave vector is near $\v C$, rather than $\v G$ in view of scatterings between electron pockets. This is because $S_{\rm SDW}(\v C)$ has already been enhanced at higher energy scales and remains to be the leading eigenvalue. To see the hidden second leading features, we show in the inset $S_{\rm SDW}(\v q)$ versus $\v q$ at the final stage of the flow. The peak near the corner corresponds to $\v C$. There is also a flat bottom near the center, with weaker strength. Inspection of the form factor reveals that it corresponds to a checkerboard spin structure, thus it actually describes $\v G$-SDW interactions. Attractive pairing interactions are induced after the intermediate stage, as seen in the flow of $S_{\rm SC}$ shown in Fig.\ref{caseA}(d). It eventually diverges so the system will develop SC below the divergence energy scale. From the inset we see the associated pairing gap in the band basis (color scale) changes sign across the two intersecting elliptical pockets. This is the nodeless $d$-wave pairing if viewed in the large Brillouine zone (LBZ). Since a strong $S_{\rm SDW}(\v q)$ would require the singlet pairing gap to change sign at two momenta connected by $\v q$, the above result seems to be counter-intuitive since $\v C$-SDW is strongest but it can only connect electron pockets to the sunk hole pockets. To have a better idea, we look into the pairing function in orbital basis to find $\phi_{\rm SC}(\v k)\sim [\alpha+\beta(\cos k_x+\cos k_y)]\tau_3 s_3$ where $(\al,\bt)=(0.22,0.44)$, $\tau_3$ ($s_3$) is the third Pauli matrix acting on the two orbitals (sublattices). The $\beta$-term describes pairing on like-sublattice, or second-neighbor bonds, and the $d$-wave character comes solely from the orbital~\cite{YWanEPL} and sublattice structure of the pairing function. Thus it takes advantage of the strong $\v C$-SDW for the pairing bonds, while the $d$-wave sign structure via orbital pairing is compatible to the weaker $\v G$-SDW which does connect the two electron pockets. This observation also shows that the $\v G$-SDW is only a balance tipper rather than the driving force. Indeed, although the hole pockets lie below the Fermi level, the $S^{\pm}_{eh}$ pair scattering works in the intermediate energy window, and via Cooperon excitations on the hole pockets can lead to attractive pair scattering on electron pockets at lower energy scales.

\begin{figure}
\includegraphics[width=0.5\textwidth]{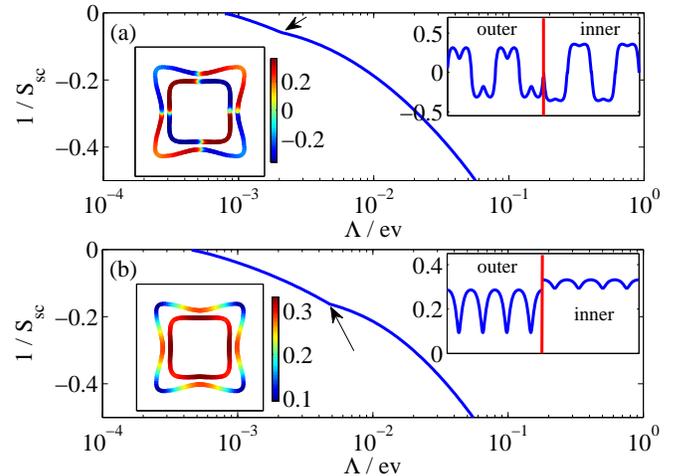}
\caption{SC flows and pairing symmetries for (a) $t_{\perp}=0.02$eV and (b) $t_{\perp}=0.1$eV. The arrows indicate the level crossings. See the text for details. The left insets show the pairing gap projected on the two electron pockets. The right insets are the corresponding angular dependence on the inner and outer pockets. In (a) the pocket splitting is exaggerated for a better view.}\label{caseBC}
\end{figure}

As argued by Mazin, the nodeless $d$-wave must develop nodes once the electron pockets are hybridized~\cite{KFSHybrid}. The question is whether it would be immediately unstable and yield to other pairing symmetries. To answer this question, we set $t_\perp=0.02$eV and perform SMFRG for the coupled layers. The flow of $S_{\rm SDW}$ is quite similar to the case of $t_\perp=0$ (thus not shown), as the $\v C$-SDW has been built up already at higher scales. On the other hand, it is still $S_{\rm SC}$ that diverges at low energy scales, as shown in Fig.\ref{caseBC}(a). There is no inter-layer pairing. The cusp highlighted by the arrow in the main panel denotes a weak level crossing of the leading pairing function. The intra-layer pairings on the two layers change from being anti-phase to in-phase. (In the case of $t_\perp=0$ the two patterns are exactly degenerate.) Apart from the layer-wise phase locking, the eventual pairing function is similar to the case of $t_\perp=0$, suggesting that the pairing mechanism is also identical. The left inset shows that on the Fermi surface the gap function (color scale) now develop sign changes, revealing $d$-wave symmetry on each split pocket. The sign change appears to be rapid, and is more clearly seen in the angular dependence of the gap function on the inner and outer pockets shown in the right inset. Thus a nodal $d$-wave pairing is realized.  Notice that there are four electron pockets according to our choice of unit-cell, but the other two are very close, respectively, to that presented in the inset of Fig.\ref{caseBC}(a), with similar sign structures. [The situation is the same in the next case study.]

We now address a situation more relevant to experiment. At $t_\perp=0.1$eV, the pockets shown in the inset of Fig.\ref{caseBC}(b) are split in momentum space to a similar extent in experiments~\cite{HDingKFeSeEPL,HDingKFSPRL}. The flow of $S_{\rm SDW}$ is still similar to the case of $t_\perp=0$. However, from Fig.\ref{caseBC}(b) we see level crossings from $d$-wave to $s$-wave symmetry (denoted by the arrow) in the flow of $S_{\rm SC}$ before it diverges. The pairing function on the Fermi surfaces (insets) is now nodeless, and it does not change sign across the split pockets. In other words, this is a $S^{++}_{ee}$-wave pairing. The pairing form factor in the orbital basis now changes to $\phi_{\rm SC}(\v k)\sim \al+\bt(\cos k_x+\cos k_y)$ with $(\al,\bt)=(0.19,0.30)$. The pairing has an onsite part and a stronger part on second-neighbor bonds, thus again taking advantage of the $\v C$-SDW fluctuations. Both parts are diagonal in layer, orbital and sublattice bases. Apart from quantitative difference in the coefficients $\al$ and $\bt$, this is in fact identical to the case in iron pnictides where both electron and hole pockets are present. If we project the gap function on the sunk hole pockets the sign is different to that on the electron pockets. Therefore $S^{++}_{ee}$ is actually a hidden, or a remnant of $S^{\pm}_{eh}$. The reason is quite simple. As long as the repulsive pair scattering from $\v G$-SDW fluctuations does not overcome the attractive one arising from Cooperon excitations on the sunk hole pockets (via the $\v C$-SDW fluctuations), the pairing gap has to be in-phase in all electron pockets. We find the same pairing symmetry for $t_\perp\in [0.04,0.1]$eV, a range that should cover the experimental situations. On the other hand, we also tune the interaction parameters to find that results are qualitatively robust apart from numerical changes in the critical scales.

Thus the above systematics identifies the $S_{ee}^{++}$ pairing (with
inphase pairing gap functions on the electron pockets) in iron selenide superconductors, and brings about a pairing mechanism involving two competing driving forces, namely, the attractive pair scattering mediated by Cooperon excitations on (virtual) hole pockets and the repulsive pair scattering from G-type spin fluctuations. The former is made possible by the surviving strong C-type SDW fluctuations and the latter is weakened by hybridization splitting. We now address experimental perspectives in view of the $S^{++}_{ee}$ pairing.

\begin{figure}
\includegraphics[width=0.5\textwidth]{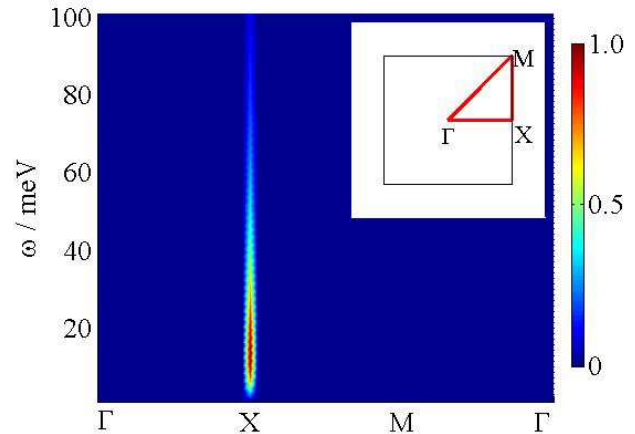}
\caption{Spin spectra (imaginary part of the spin susceptibility) in the $S^{++}_{ee}$ SC state along a line cut (in LBZ) shown in the inset. The data is rescaled by the maximum value.}\label{neutron}
\end{figure}

Regarding the neutron experiment, the usual wisdom is a neutron resonance occurs if the gap function changes sign. However, given the fact that $S^{++}_{ee}$ is related to $S^{\pm}_{eh}$ and that $\v C$-SDW is strong, a neutron resonance is not impossible. In Fig.~\ref{neutron}(b) we show a model calculation of the dynamic spin susceptibility in the $S^{++}_{ee}$ superconducting phase within the random-phase-approximation, using the renormalized spin interactions (scaled by a factor of order $0.1$ to avoid divergence). The spectrum along a line cut (shown in the inset) shows that the intensity is strongest around $\v C$ at a frequency $\nu=14$meV. In fact even in the normal state the intensity is strong at the same momentum, in agreement to a recent neutron diffraction experiment~\cite{KFSINSBirgeneauPRL}. Our result is notwithstanding the experiments since, as we mentioned, the resonance momentum in the neutron data varies from $\v C$ to half way between $\v C$ and $\v G$~\cite{KFSINSInosovPRB,KFSINSInosovPRL,KFSINSInosovEPL,KFSINSDaiPRB,KFSINSBirgeneauPRL}.

We also notice that the $S^{++}_{ee}$ phase can benefit from electron-phonon coupling which causes attractive pair scattering only, and thus may have an elevated $T_c$~\cite{SMFRGFS}. Instead, the electron-phonon coupling would try to lock up the gap signs on the electron pockets and thus would be destructive to $S^{\pm}_{ee}$ pairing. Another issue is the impurity scattering. In this regard, the $S^{++}_{ee}$-wave pairing is effectively similar to a single-band $s$-wave, thus is robust against scalar impurity scattering via Anderson theorem. Instead, the $S^{\pm}_{ee}$ phase would be subject to pairing breaking effects from scalar as well as magnetic impurities, and this is difficult to reconcile the experimental situation that even dirty samples are good superconductors.

Finally we propose two further experimental probes to verify/falsify the $S^{++}_{ee}$ versus $S^{\pm}_{ee}$ pairing. In scanning tunneling microscopy, due to the coherence factor, the quasiparticle interference pattern should be enhanced (weakened) as time-reversal-symmetry is broken (e.g., by magnetic impurities or under an applied magnetic field) for $S^{++}_{ee}$ ($S^{\pm}_{ee}$) pairing. On the other hand, in the transparent limit of an out-of-plane contact junction, the Andreev reflection is enhanced (weakened) for $S^{++}_{ee}$ ($S^{\pm}_{ee}$) pairing due to quantum interference between the Andreev reflections from the two electron bands~\cite{wangda}.

While finalizing the manuscript, we become aware of two parallel works addressing the same issue~\cite{JXLi,DHLee}, where single-layered models with phenomenological off-site spin-exchange couplings are assumed.

\acknowledgements{The work was supported by NSFC 10974086 and
  10734120, the Ministry of Science and Technology of China (under
  the Grant No. 2006CB921802 and 2006CB601002) and the 111 Project
  (under the Grant No. B07026).}

\bibliography{iron_smfrg}

\end{document}